# Research in the Resources Framework: Changing environments, consistent exploration

Michael C. Wittmann
*Department of Physics and Astronomy, Maine Center for Research in STEM Education,*
*University of Maine, Orono ME 04469*

In this paper, I discuss my personal journey through one research tradition, that of the resources framework, and how it has evolved over time. In my present work, understanding learners' reasoning in physics in terms of the construction of large-scale models from small-scale resources emphasizes the person doing the constructing over the physics they are discussing. In this human-centered approach, I find value not in the correctness or incorrectness of a given response, but in the nature of construction, the individual's evaluation of their own ideas, and the communication between learners as they seek to understand each other. The resources framework has driven my attention toward a human-centered approach, and has had an effect on both my professional and personal life, in the process. In addition, events in my personal life have proven relevant to my professional work in ways that are reflected by my use of the resources framework to understand knowledge and learning.

## I. INTRODUCTION

In studying the teaching and learning of physics, I have over time paid more attention to the person doing the learning, rather than the material that they are learning. This attention to the individual and their community was barely visible at the start of my career, but has grown stronger over time, affecting my methods for gathering, analyzing, and reporting on data. In some ways, I have moved from a clinical approach involving a kind of distant observation to an ethnographic approach of embedded observation, with me within the communities that I am studying. Throughout, though, I have used the resources framework to make sense of my observations. The use of resources has helped me in this transition. The framework has pushed me to emphasize the learner as they are, rather than the physics as it should be. To illustrate these changes, I describe my work at three times in my career using papers that use and investigate the resources framework in a different way.

To begin, though, I summarize the resources framework. I provide several examples of resources and how they simplify reasoning about physics problems. I end this section by raising three research questions that have played a role in my work within the resources framework. In the second section, I present a general outline of my research activity, summarizing in broad brush strokes the activities I was engaged in, and how the individual papers were but elements of larger interests at the time that they were written.

## II. A SUMMARY OF THE RESOURCES FRAMEWORK

In this section, I summarize the resources framework and situate it in a historical and developmental context before summarizing some of the ways that it has played a role in the PER community.

A resource, as originally defined by David Hammer [1], is an isolated, independent, productive idea that plays some role in solving a problem. An example of a resource is "part-for-whole," where a part of an object or system can represent the whole of it [2]. In physics, we use "part-for-whole" when we represent the motion of an object by the motion of its center of mass or some other salient feature. Our thinking is simplified by representing a large, complicated system by only a single part of the system. For example, the center of mass of an object travels a parabolic trajectory, even as the object rotates. The idea can be misapplied, too: for some, the peaks of waves are the only parts that superpose [3]. The idea of part-for-whole is used in other contexts, as well. In politics, a president or queen represents a country. In corporate organizations, a charismatic Chief Executive Officer - CEO - represents a company. In literature, the concept is so common that the term synecdoche is used to describe it.

A separate example, "dying away," (first discussed by diSessa [4]) illustrates some of the complexity of understanding and investigating the use of resources. The resource might describe the motion of a box pushed across a frictional floor, the ringing of a bell, or the amplitude of a propagating wave [5]. By using a single resource, it's possible to describe a complex event without providing more detail – a ringing bell fades out for complex reasons, but it always fades out over time and sometimes that is all one cares about. Still, the resource of "dying away" contains substantial structure. There must be a property of an object, and this property must decrease over time. To think about objects, properties, and time is anything but simple, but to the user of the idea, the resource can simply be applied without further reflection. To a researcher, though, it might be valuable to think about how a resource came to be, perhaps by using models such as the reification of processes [6–8] or conceptual blending [9].

The idea of resources was developed after diSessa's description of phenomenological primitives (p-prims) [10,4,11,12]. Scherr summarized the essential elements of the resources framework in a paper which distinguishes between a "misconceptions" and a "pieces" (*e.g.*, p-prim or resources) approach to modeling student reasoning [13]. In the resources framework, a particular resource need not be correct or incorrect in and of itself. Each resource has some value when used, and can be thought of as lying in wait, ready to be used when cued. Sometimes, the wrong resource might be used, or a perfectly useful resource might be used in the wrong situation (e.g., "dying away" might be applied to the impetus of a ball thrown in the air). The use of a resource may be problematic, but the resource must be useful at least in some settings, otherwise a person would not have developed it for their own use. Many resources can be and usually are applied to address a single problem, but resources may or may not be used with each other coherently, allowing for contradictions in what student responses to our questions. Resource use is highly context dependent, as determined by the student's views of context, rather than an expert's. To a researcher, investigating resource use would help understand how different populations engage with similar contexts differently, or how populations engage with different contexts differently, especially contexts that to a physicist are the same but to a student are not.

A resource, as originally defined, is a chunk of knowledge that can be used in a particular problem solving setting. This idea of "chunking" knowledge is studied in the world of psychology, with the well-known result that people's working memory can make use of roughly 5 to 9 "chunks" of information at a time [14]. Typically, many resources must be used together to solve a problem. Within the resources framework, learning can then be described as knowledge construction, with resources as building blocks that come together to form more complex ideas. diSessa and Sherin discussed networks of knowledge pieces and how they are activated based on one's readout of a given situation through the further construct of coordination classes [15]. Sabella and collaborators described the ways in which different cues can influence which elements of overlapping networks of resources get chosen [16]. In previous work, I applied the resources framework to four kinds of conceptual change [17]. To a researcher in the resources tradition, investigations of learning might involve studying the construction of resource networks, their judicious cuing in situations, and their reorganization by a variety of methods.

The resources framework has been extended and applied extensively. These studies have created a formalism and provided case studies to help understand knowledge and learning within the framework. diSessa described the use of knowledge pieces in terms of the span and alignment of p-prims [4] within and across contexts. Hammer et al. discussed the cuing and activation of resources in the context of framing and transfer [18]. Redish connected the resources framework to what is known of neurocognitive modeling of the mind [19]. To a researcher, the creation, choice, networking, and use of a resource within a given problem can all be of interest, particularly when the answer depends on nuances of problems that were not originally noticed by the researcher.

Resources, as originally defined, were conceptual in nature. Hammer and Elby extended the model to issues of personal epistemology [20–23] and its role in the classroom. They applied the framework to the questions "how do I know?" and "why do I believe this?" An epistemological resource might be "someone told me" or "I figured it out." Like with conceptual resources, neither epistemological resource is wrong, but one might be more appropriate in a given setting than another. More to the point, one might be more productive, helping the person doing the thinking engage more usefully with a given problem. To a researcher studying the complexity of problem solving, there might be value in considering other kinds of resources, beyond conceptual and epistemological, that help make sense of problem solving.

Other models with similar goals for research and teaching illustrate one further area of research within the resources framework. In Minstrell and collaborators' models of facets of knowledge [24] and the BOLT framework [25], the term facets already suggests one of the issues with resources. Earlier, I said a resource is an idea used for problem solving, and later I said that a resource could contain substantial substructure. I also noted that networks of resources might be used together, and some might reify into commonly used, new resources (which might be unpacked to show the substantial substructure of that resource, if needed). In case that isn't complex enough, one can also think of there being different facets of a given resource. In one context, a resource might be applied one way, and so regularly that it seems like its own idea. In another context, the same resource might be applied another way, and so regularly that it seems like a different idea. Modeling this connection of resource and context can be very productive (see the Diagnoser project for more details [26–28]), but is a different approach than looking at the resource by itself. To a researcher, explaining the interplay of a resource and its facets might lead to questions about how to evaluate the usefulness and productivity of the resource in different contexts.

The resources framework comes with certain built in assumptions and methodologies, in part derived from how it was developed. Hammer introduced the term in 2000, building on work done previously by diSessa, which was first published in 1983 [10]. diSessa's work, in turn, arose in part through his interactions with Seymour Papert, a former student of Piaget's, on the computer language Logo and its application to mathematics learning, Turtle Geometry. Papert had previously collaborated at MIT with Marvin Minsky, who talked about agents in the mind [29]. In other words, the resources framework comes from a long tradition, stretching back at least to the 1960s, of investigation into small-scale ideas that are connected to each other in order to address large scale problems. It is worth noting that other models, such as intuitive rules [30] and symbolic forms in mathematics [31], have also been used to model small-scale reasoning (rather than large-scale concepts) about a topic. Importantly, these approaches have focused primarily on the individual, and have often gathered data in situations where an expert is talking to a novice (be it in clinical interviews or classroom interactions), though exceptions exist. An open question in studies within the resources tradition has been to see if and how these ideas about the mind can be applied to group learning situations and interactions between people.

The papers I discuss below address several of the questions raised in this section. First is the question of how we observe resources, whether through written or multiple-choice questions, interviews, or in group learning situations. Second is the question of what kinds of resources exist. Hammer and colleagues proposed conceptual [1] and epistemological resources [20–23], and we suggested there might be others, particularly in the area of mathematical problem solving. Third, given the complications of the resources framework, making decisions about which resources to pay attention to, as instructors and facilitators of learning, requires an understanding of the conflicts between useful and problematic facets of resources in a given situation. I have summarized these issues as well as others that arise in this paper in Figure 1. The arrows in the figure indicate that there was development along each of these axes, as will be described in the remainder of this paper.

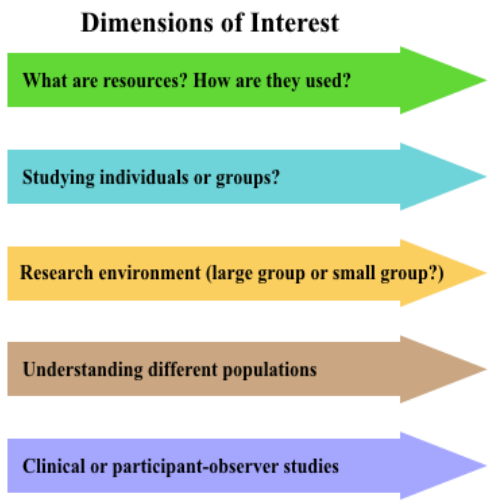

*Figure 1.* Dimensions of interest in carrying out studies of the resources framework (a non-exhaustive list)

## III. THE RESOURCES FRAMEWORK WITHIN A BROADER RESEARCH HISTORY

To show how questions within the resources framework, and associated questions about the limits of the framework, have affected my professional work, and to place the three papers I describe in this paper in context, I present a brief summary of my work within physics education research.

My training in physics education research began in the mid-1990s within the tradition of investigating specific student difficulties and addressing them by creating new targeted learning materials [32,33]. At the time I did this work, I didn't think too much about its larger structure. Like many novices, I specialized in what was in front of me, and only later broadened my attention enough to see other approaches. In my case, I didn't think much about what a "specific difficulty" was, in part because I didn't know to ask, though others were asking and answered the question later on [34]. Only later did I think more about the complex interaction of content knowledge, classroom materials or assessments, and the goals of instruction that made a "specific" difficulty much more than simply a cognitive, conceptual problem or a misconception.

Instead of noticing these issues early on, I was more interested in the cognitive modeling being done with knowledge-in-pieces approaches, as defined by diSessa's work [4,10]. In part, this arose from my data, where I kept finding that student responses to questions about mechanical waves sounded like students thinking about material objects being thrown or bouncing off each other; I was able to describe these answers in detail using diSessa's carefully described p-prims about forces and motion [4]. I became a passionate cognitivist during this period, modeling how people think in terms of building blocks of their reasoning. Given that I was E.F. "Joe" Redish's student, it's no surprise that I was strongly influenced by his 1994 paper on applying cognitive theory to teaching physics [35]. David Hammer arrived at the University of Maryland just as I started my post doc. His work on knowledge-in-pieces was relatively recently published [36], and his mentorship around research, teaching, and how teachers engage with students only strengthened my involvement in what would be called the resources framework. I focused more on the modeling of content knowledge than on the people doing the modeling, though. I was not yet paying attention to what Eleanor Duckworth called the "having of wonderful ideas" [37]. But, the framework provided a pathway for me to change. I liked the way the research model connected to the craft of teaching through facilitation, helping people use their own good ideas to help them learn.

At the time, funding came through curriculum development. In the years of my PhD, post doc, and early faculty career, I developed classroom materials based on both specific difficulties and productive resources, trying to reconcile the two approaches [38,39]. My approach to developing new learning materials was eventually guided by a kind of activity theory interpretation, where the materials, the goals of instruction, and the interactions of instructors and students each had an agenda that needed to be balanced in the moment of instruction. I used curriculum development to continue my work on the resources framework, with excellent collaborators throughout the years [17,40–46].

In evaluating the curricula I was developing, and because the classes I was studying were so small, my research methods changed to include more than written questions and interviews. Guided by questions about what happened during instruction, rather than before and after as measured by surveys or written questions, I had begun to study video of classroom interactions. This had begun in the late 90s while studying quantum physics [47], and continued with investigations of mechanics and teaching of quantum physics to non-science students. In the process of studying student interactions, my empirical toolbox expanded to include gesture analysis [48,49], discourse analysis [50–52], and embodied cognition and interaction [53–56]. Using a different toolbox to observe student problem solving led to new work within the resources framework. Rather than asking only about what knowledge students have, we also spent time looking at what they were doing, and how that intersected with their knowledge. This focused my attention on the learner more than their ideas. Stated differently, I became less interested in the nature of the resource itself and began to look more at how people applied them. I shifted from studying the nature of physics knowledge to studying the nature of people's use of their knowledge.

In 2010, an unplanned opportunity arose to work not with college students but with middle school teachers. Though I had worked with pre-service teachers previously [57,58], I was suddenly working with in-service teachers. Rather than investigating student ideas, I was investigating the complexity of teacher pedagogical content knowledge [59–63], looking primarily at teacher content knowledge and how teachers thought about student ideas. This work was guided in part by the group who developed Cognitively Guided Instruction [64–67]. I continued to study particular content areas like energy [51,52,56,68] or accelerated motion [69], and to develop theory within the resources framework and knowledge-in-pieces tradition [70–72]. I continued to investigate student knowledge [51,52,73], though other studies focused on teacher knowledge [74] in a way that could feed back into how we interact with teachers [69,75–78]. But, my concerns were now deeper than just understanding the knowledge teachers had. I was also invested in understanding how they used that knowledge in their teaching.

The work was now community-centered, teacher-oriented, and practitioner-focused, and I was but one member of a large and complex community. What I brought to this community was, in part, my emphasis on the resources framework, a theoretical approach that had been shown to lend itself to making sense of student ideas, and perhaps helping the teachers think about different ways of being more responsive to the good ideas students brought to the classroom.

# IV. THREE STUDIES USING THE RESOURCES FRAMEWORK

In this section, I use three papers written roughly ten years apart to show the at times unspoken ideas about resources that guided my work. The first paper had the most clinical focus while investigating students in an introductory physics course. There was a definite attention to correct and incorrect ideas, and little attention to the people having those ideas. The second paper looked at student-student interactions during an intermediate mechanics course, but still with a clinical focus. Importantly, the two student solutions discussed in this paper were both correct methods, and the attention was on the two students as much as on their ideas. The third paper looked at a large-group discussion during a professional development (PD) meeting with in-service teachers. The research question only emerged after the meeting, but was deeply informed by my experience within the resources framework. I note that I regularly use the singular when writing about these papers in this section because I want to emphasize that these are my thoughts, and not those of my co-authors.

## A. Comparing student reasoning on written and multiple choice questions

In 1999, together with my post doc mentor, Richard Steinberg, and my PhD advisor, E.F. "Joe" Redish, I published a paper in The Physics Teacher [3] about student ways of thinking about the propagation and superposition of mechanical waves. The overall claim of our paper was that students were using ideas that could be used productively when describing particles or objects, and mis-applying these ideas in the context of waves. At the time, Hammer had not yet published the name "resources" but I was already using diSessa's ideas about knowledge pieces and p-prims in my dissertation work. My goal in this section is not to revisit the paper in detail, but to give insight into the reasoning behind the paper, especially in contrast to the following two papers.

When studying propagation, we were comparing the use of two different questions. One was a free response question, in which students were given a picture of a wave-pulse traveling on a long, taut string, held by a hand (shown in a figure provided in the question), and asked to say how one might speed up the wave. A variation on the question asked for a wave-pulse that reached the wall sooner, emphasizing time rather than speed. The second was a different kind of question, which we called multiple-choice, multiple response (MCMR), in which students had to answer the same question, but we gave them a long list of possible answers, and they had to check off all that apply.[1]

During the 1980s and 90s, diagnostic surveys like the Force Concept Inventory were being developed and having a large impact on physics teaching [79,80]. Research on student misconceptions was popular, and instructors in workshops would ask for lists of misconceptions, so that they could be aware of what their students were thinking incorrectly. In this culture, evidence of good teaching was often provided using a question on which most students avoided the purposefully provided and research-based incorrect distractors and chose the correct item on a multiple-choice survey, instead. A good question had research-based distractors that were tempting to students; statistics on these surveys were often worse than guessing as a result.

---

[1] These questions are sometimes called multiple-true/false questions, since each suggested item is actually a true/false statement.

At Maryland, we were developing a concept inventory on mechanical waves. We wanted to write multiple-choice questions that allowed quick and easy analysis of student responses. Our hope was to come up with ways to ask multiple-choice questions that would let us observe the knowledge pieces that students had. In this context, a free response question might be the worst kind of question, because students would answer with something that was sufficient, but not necessarily complete. The multiple-choice, multiple-response (MCMR) question in the 1999 Physics Teacher paper was written so that it would hopefully trigger *all* the ideas that we had seen students bring to this question in our interviews. Through such a question, we might be able to gather rich data like from free-wheeling interviews that could explore ideas from many perspectives, but quickly, easily, and in a way that could be quickly interpreted. From a resources perspective, the MCMR question was designed to elicit all the knowledge pieces that students might bring to the problem.

In parallel to this development of a concept inventory, we were engaged in research-based curriculum development. We were developing a series of small group learning activities, tutorials, to complement the University of Washington-style tutorials that we had been piloting at the University of Maryland for several years. I'd been a teaching assistant and head teaching assistant for the implementation of these materials, had taught hundreds of hours of tutorials, had helped others develop the skills to do the same, and was very much interested in the artform of asking the right question, at the right time, to elicit student ideas in just such a way as to help them grasp whatever idea was just around the corner (or on the next page).

Though I had started in what came to be called the "specific difficulties" school of thought, I had also become aware that the knowledge-in-pieces model was productive in helping me understand what I was hearing students say about mechanical waves. Student responses to a whole series of seemingly unrelated questions about mechanical waves showed evidence of thinking using resources commonly associated with objects, not waves. Students seemed to interpret the peak of a wave as if it was the center-of-mass of an object. If we thought of their answers as if they were talking about objects, and not waves, many of their responses began to make sense.

Students would often talk about "dying away" when talking about the mathematical formalism of wave propagation [5] and the way that the amplitude of a propagating wave would die away. Similarly, when discussing sound waves propagating through air, students would use the idea of "maintaining agency" (a force is required to keep things moving, for example [36]), refined to include concepts of frequency and volume and their effect on the initial force on the system [81].

In the 1999 paper, we discussed several resources but did not discuss them as resources. Instead, we summarized student thinking in terms of mixture of ideas from how particles behave and how waves behave (see Table III in [3]). Our description that "a harder flick of the wrist implies a faster wavepulse" was a facet of Hammer's "actuating agency" [36] and diSessa's "force as a mover" [4]. The statement that "smaller pulses can be created that move faster" was not directly connected to a previously described knowledge piece, but could be thought of as a facet of Ohm's p-prim [10]. When we wrote "Only the peak of the wavepulse is considered when describing superposition," we were referring to what we eventually came to call the part-for-whole resource. When stating that "wavepulses collide with each other and they cancel or bounce off each other," we were referring to diSessa's description of the "bouncing" and "canceling" p-prims [4].

That we did not describe our results in terms of resources was a mistake, from my present perspective, because it hid the intellectual work we were doing. We were using the terminology in other settings, but not publishing them here. Though they helped make analysis and categorization simpler, we were writing for an audience in which we did not think the technical language would be of help. This undermined some of the value of the work, because readers would not be able to make the connections to the literature that had guided our own thinking. Still, the work had other meaningful elements that connected to research done in the resources tradition. For example, the study of multiple question formats was found to be helpful for tapping into more student ideas than might be shown in a free response question.

### B. Defining procedural resources from group interactions

By 2007, roughly a decade after the work described in the Physics Teacher paper, I was a faculty member who had taught many smaller (10 to 20 person) upper division courses. I was not dealing with introductory topics, many of which could be described using knowledge pieces that had been developed as primitives much earlier in life. In these upper division courses, calculus, vector calculus, and differential equations served to represent the conceptual language of the course, and students seemed to be developing a kind of fluency with the mathematics that showed they were turning some of these mathematical ideas into resources for more complex problem solving. I wanted to understand the use of mathematics in upper division physics and to model this understanding using a resources framework. Other groups (in Colorado, Maryland, and Washington, for example) were also studying the role of mathematics in upper division physics, and I thought my focus on a resources framework would set me apart from their efforts.

The small classes I taught forced a change in methods for gathering data. Surveys would be relatively meaningless, since statistics for small classes would contain too much variability. Instead, building on methods I first implemented while a post doc at Maryland and influenced by David Hammer, we filmed the classrooms and carried out group interviews. A video camera provided an easy way to gather information about people as they did their work, and interactions in group interviews mimicked that behavior in a more controlled setting. The methodological shift to video observation forced a profound change in what I attended to in student knowledge. I was not looking at pre- and post-instruction surveys, seeking changes in knowledge over the long time scale of a semester, and therefore was not as focused on crafting questions that would elicit a particular piece of knowledge. I was looking instead at what students were saying in the moment. The focus was still cognitive, but the attention was now on the individual student, not a collection of students. The ensuing changes in my research are shown in a 2007 paper with Katrina Black [42], where we looked at problem solving in a sophomore level mechanics class through the lens of the resources framework.

An additional point about my work arose from my non-work life. My home life aligned nicely with these new research methodologies. I was spending a lot of time simply watching my children (who were 3 and 6 in 2007) and figuring out how they were making sense of the world. When someone has only seen 2 winters, and probably has no memory of them, how do they respond to this utterly weird fact of the weather getting so much colder?! Parenting was endlessly fascinating to me, in particular with its focus on this particular child doing this particular thing at this particular moment. I was living a case study mindset at home, especially once I had two kids who were so very different from day one, and it's no surprise the same would happen at work.

An important element of this was that I was raising my children bilingually, meaning that we were regularly speaking two languages at home. The kids spoke one language with one parent and another language with the other. If they used the non-favored language with a parent, the parent could shift into that language with them, insist on keeping to the favored parental language, or skip back and forth. It was a very resource-y approach to language. There was no wrong language, and there were always multiple correct ways to say the same thing (that delicate flapping thing with the symmetric patterns is both a butterfly and a Schmetterling). This aspect of duality in my home life came to influence my work life, as described below.

Funding for my work came primarily through curriculum development. In the process of developing the Intermediate Mechanics Tutorials [39,82], I was focused on combining the use of mathematics and physics, as I had done previously in the Activity-Based Tutorials [33]. My collaborators, Ellie Sayre and Katrina Black, played an enormous role in the research and development of these materials. In studying intermediate and advanced physics courses, I would often say that the mathematics was not separate from the physics, as it might seem when distinguishing between conceptual understanding and problem solving in introductory physics. Instead, the mathematics counted as the conceptual shorthand for the advanced physics. To carry out a mathematical procedure was to do conceptual work within the physics. When modeled in terms of resources, this meant that mathematical procedures were a kind of resource, different from the conceptual resources Hammer had described, but equally useful for problem solving. Hammer and Elby had used a similar argument to posit the existence of epistemological resources [20–23].

Our original goal was to understand how new procedural resources came to be. By this point, I had developed an activity theory perspective on specific student difficulties as the intersection of student knowledge, curriculum goals, and contextual cues, and I had developed a resource-based perspective on misconceptions as highly coherent, consistently constructed and reified sets of resources. Originally, we were focused on studying the process of reification for specialized mathematical resource like "integrate," for example. Eventually, a student might just say that you integrate an equation and carry out the task, but early in the learning process, they might have needed to think about each step of the procedure, and the full act of integration as a long chain of procedures. Much like in Sfard's process-object theory [8], from mathematics, students might eventually make a single conceptual object, "integrate," out of a connected set of procedures. The problem was that this kind of developmental perspective was not available to us, given the lack of longitudinal data we had for students in our class. Instead, we were observing them at very select times, in great detail, and for a short amount of time.

Because we were studying video, I was paying far more attention to the people doing the thinking than I had done in the 1999 study. In our 2007 paper, Black and I used transcript from a single conversation between two students to look at the resources each of the students was using, and why they were at times talking past each other, even while using the same resources. The task involved solving an integral with some under-specified initial conditions, observing how students brought physical meaning to their integral solutions by determining the values of the undetermined constants in their work. One student was using limits of integration to solve the problem, and the other was using undetermined constants. Both methods were correct and appropriate. The students were using some of the same procedures, such as "compute anti-derivative," a complex act in and of itself. They were also using different procedures. One used "add constant," but did not seek to "find value" for the constant in the way that we observed other students using the undetermined constants ("+C") method did. The other student "chose limits" after "extracting boundary conditions" from the problem statement, and "applied limits" when taking the anti-derivative. The former student solved the mathematical problem using +C, but without embedding the answer in the specific physical situation. The latter student used integration limits to give a complete physical picture in which the boundary and initial conditions were taken into account. In work that followed this 2007 paper, we regularly observed that students using either integration method used "compute anti-derivative" and "extract boundary conditions," but we observed different resources in what they did with the boundary conditions, either "add constant" and "find value" or "choose limits" and "apply limits." In the case of the 2007 paper, the mixture of shared resources and different resources created problems for the students as they talked to each other. Driven in part by my attention to my children's bilingual lives, I was ready to explore this idea when Katrina Black brought it to me. It also helped strengthen my understanding of Sabella's work on problem solving using overlapping networks of resources [16].

We learned three things from our analysis. First, as a proof of concept, we felt we had shown that there were procedural resources, and that it was useful to think of them like other resources: independently cued, connected into complex networks, and used to solve some kind of problem. Second, we were able to observe different scales of resources for a single task. What was a reified single procedure for one student ("you just integrate") might be a complex map of procedures, for another student ("first you do X, then you do Y, but for that you need Z") [8]. Pedagogically, this helped us unpack a given idea into its constituent parts, with consequences for how to structure instruction and listen to students. Finally, by modeling the situation to show that the same procedural resources might be part of two different solution paths for students, we were able to explain some of the confusions that arose in student interactions, where students might not recognize that some procedures, like "add constant" weren't useful when using the limits of integration method. We had, it should be noted, observed students using both the "+C" and the "limits" procedures at the same time, expending much effort to find that the undetermined constant had a value of 0. Arriving at these findings had required changes in research methods brought on by changes in class size, acceptance that our original research goal was not feasible using our research tools, and the discovery of results that resonated with a complex interaction of personal and professional interests.

I now recognize that I was involved in a different kind of work than in 1999. We developed and extended the resources framework to include procedural resources in addition to conceptual and epistemological resources. To do so, though, we paid attention to a conversation between two people. We had to pay attention to conversation and discourse, and we had to listen for the particular details of an individual student in the process. Given the highly variable nature of resource use, students will create their own unique constructions to solve a problem. To attend to their construction required attending to both their ideas and their unique construction. This in turn led to more attention to the person having the ideas, and not just the ideas themselves.

### C. Teacher noticing of problematic facets of a productive resource

In the third section of this paper, I describe a situation in which the resources framework was no longer the primary agenda of the research, but instead our work on it emerged over time as we made sense of our observations.

The work took place within a professional development environment, working with in-service teachers on the teaching and learning of energy in middle school science as part of the Maine Physical Sciences Partnership. Within this work, the in-service middle school teachers were partners, rather than subjects of my research. My work with teachers was around their content knowledge, knowledge of students' ideas, and ways of responding to these ideas. I had quickly learned that teachers' motivations for responding to students were only partially connected to the correctness of a given idea. They were also often guided by issues around inclusiveness, attitudes toward science, promotion of curiosity and inquiry, and many other topics that are often treated as a hidden agenda in college classes. Though teachers did not use the language of the resources framework in their discussions of teaching, teachers were naturally engaged in resource-like thinking as they weighed decisions about how and what to respond to in the classroom, and it was a common theme of our conversations. Specifically, they recognized that students' answers had many different elements, that these might be contradictory, and that they were often constructed on the fly, created for the first time in that moment in that classroom. Teachers tried to value the good things that students said as a way to keep the students motivated and engaged. In sum, ideas very much consistent with the resources framework were part of their everyday classroom work.

This thinking on the teachers' part had a major influence on my thinking as I worked with and learned from them. Because of my past work with classroom video, my perspective had become more ethnographic and attentive to community as a whole, not just the knowledge within that community. Supporting and sustaining the community was more important than gathering any particular data, and leading a professional development activity that provided value to teachers took precedent over any kind of research goal for a given task.

This is little different from doing research in one's own class, of course, but it felt different in one particular way. Previously, such as in the 2007 study, I had on principle not discussed data my research collaborators gathered in my classes during the semesters I was teaching those classes. I wanted to avoid the conflict of interest inherent in knowing more about the students than was in the activities that went into the gradebook. With the teachers, this agenda changed in part because my position in the community was different. Except for my physics knowledge, I was the least knowledgeable person in the room when it came to understanding middle school teaching, school systems, and the like. Other than my status and position as a university researcher, I had little status, yet was often leading activities with teachers. In sum, we were different kinds of experts. I had never seen students as my equals in the way that my teacher colleagues were now my equals, and often mentors. (I now teach differently as a result of this insight.) Further, gathering data in a way that might impact the safety and persistence of the community was unacceptable to me, not only in a formal sense according to human subjects review, but in the community sense of supporting the long-term existence of our group.

In thinking about teacher knowledge, I was using Ball's framework for understanding pedagogical content knowledge [59–62]. Because of my background, I thought of teacher knowledge in terms of resources. We had been strongly influenced in our PD design by the Cognitively Guided Instruction community [64–67], but their focus had been primarily on elementary school student knowledge mathematics. We were interested in students' and teachers' both, especially in situations where incorrect answers contained in them a multitude of good ideas and structure about learning and modeling. Combining these different mindsets meant, for example, that we wanted to see what teachers did when they found value in students' answers, regardless of these being correct or incorrect by some grading metric.

In 2017, Carolina Alvarado, Laura Millay, and I published a paper [78] (an extension of [76]) in which we discussed how a resources perspective could influence the decisions teachers make as they address student thinking in the classroom. Rather than being cognitive in focus, this work was focused on teachers being responsive to students in the classroom, in whatever way that was most appropriate [83,84]. In the process, though, we had to investigate and understand the "grain size" of a resource. Resources and knowledge pieces are often called "small-scale," in comparison to "large-scale" conceptions. In our work, we realized that there might be tensions between different correct grain-size interpretations of a given response. Though we had not entered into the work planning to do theory development, the end result showed how empirical evidence can drive theory.

In one task, we had asked teachers to think about how students might answer a survey question about potential energy, with three identical hikers taking different paths from the same starting in the valley to the same ending point on a mountain peak [85]. The prompt given to the teachers was to address why students might answer that the hiker who had taken the shortest, steepest path (answer A to the question, listed only as "Hiker 1") would have the most potential energy upon arriving at the top of the mountain. Data came from an analysis of the discussion that followed this prompt.

Methodologically, this study's questions emerged from the data in a way that was different from the two studies described previously. In both of the 1999 and 2007 studies, the research study and the questions within the resources framework had been designed into the work. In this third study, the analysis of data only slowly led to an emphasis on the resources being used by the teachers. This methodological shift in analysis was consistent with an overall move away from designed research toward naturalistic observation of events.

In part, this shift in methodology arose due to the conflict of interest that would arise, were the research goals to be more clearly followed. As facilitator of the professional development activity and as a researcher, curious about the outcomes of the activity, I acted such that my responsibility to the teachers was my first and foremost goal. Were I to focus on a particular research question, I might have ended up treating the situation like an interview, with specific probing questions meant to address some particular question. By focusing on the teachers and the goals of the professional development, though, I could emphasize the interactive nature of the conversation, concern myself with the content focus that was the publicly stated goal of the PD session, and not worry about the research until later.

That my goals of the research were closely related to my goals of the PD could be seen in the prompt given to teachers. I asked "why might a student give answer A?" so that teachers would attend to the student's construction of ideas, hoping that their responses would help us be able to talk about the good ideas students had, even when answering incorrectly. Asking "what is wrong with answer A?" would have led to a discussion of ideas that were wrong, which I wanted to avoid and which wouldn't really help teachers. How do you respond to a discussion of wrong ideas? By talking about the correct ideas. That doesn't help teachers listen to their students. Wouldn't it be better to observe what students could do, and work with that? Focusing on the students was more consistent with the teachers' overall activities in the classroom.

In reviewing the PD session, Alvarado and I quickly recognized that the teachers had provided explanations of student answers that were consistent with the research literature on student understanding of potential energy. They recognized that students might double count work and energy [86,87]. They were very much aware of the common "energy is used up" response [88–92], and gave three different versions of it. In sum, they were able to name all the typical incorrect student responses, including nuances not previously described in the literature. Upon lengthier reflection, though, we realized that a problem was lurking behind the scenes, and it was a problem that I would not have attended to were I not thinking from within the resources framework.

Many of the proposed student ideas that led to an incorrect answer were a facet of a single idea [24], namely the metaphor of energy as a substance-like quantity. This metaphor is recognized as commonly held by students [89,90,93] and researchers [94], and its importance in the creation of the concept of energy has been described, as well [95]. At the same time, some have argued that it is a problematic idea [96,97].

Within the resources framework, the energy-as-substance metaphor is a resource, and the facets of the metaphor are also each a resource. Any resource can be a useful problem-solving tool, though mis-applied in a given situation, but the resource itself is neither correct nor incorrect. What concerned us was that the teachers had only noticed the problematic application of these "facet" resources. The more general resource, the "metaphor" resource which was the basis of those "facet" resources, was not discussed as valuable. Teachers who do not notice the generative value of the "metaphor" resource during instruction might lose a teaching opportunity when discussing energy and hearing their students applying the "facet" resources of energy being used up, for example. This made clear to us how detailed a knowledge of energy was required of teachers, helping guide future professional development activities around both content understanding and how one evaluates and makes use of what others know.

Through this emergent process of analysis, we had arrived at a question about the scale-free nature of resources as described in the literature. A resource is simply a "chunk" of conceptual understanding or procedural work, to the user. To an external observer like a teacher or researcher, though, several seemingly different resources might be different contextual instances of just one conceptual "chunk." Whether teachers see conceptual knowledge in this fashion, and how they respond as a result, was a question that arose in our research, without having been the original goal of the work. Thus, our concern about the scale-free nature of resources arose in part because of the need of teachers to respond to their students.

### D. Summary

The resources framework has played a complex role in my work. I began my research with the mindset of addressing student difficulties, but used diSessa's knowledge pieces to make sense of these difficulties, while using written questions to elicit student responses. From 1999 to 2007, I moved toward attending to thinking in the moment, focused in particular on situations for which there are multiple correct answers. In my personal life, a parallel situation was occurring as I raised my daughters bilingually and observed how they negotiated the use of two languages in a variety of settings. My research and my parenting were consistent with my attention as a teacher to the individuals in my relatively small upper division classes, as opposed to the large lecture classes I had originally studied. From 2007 to 2017, I moved further into the area of naturalistic observation, focusing more on the individual and the variability of their knowledge and reasoning. Working now with a community of teachers, I observed how they paid attention to their students' ideas. Again, the resources framework was helpful in making sense of teachers finding value in what students said and did. I observed in them an attention to the student in addition to the physics idea; this strongly influenced my own approach. In sum, the resources framework has helped me focus my attention and given me insight into the people doing the thinking, even as it describes the thinking they are doing.

The resources framework and the methods used to study it drove some of the changes in my work. In terms of the theory, paying attention to the network of resources students used in solving a problem changed the locus of attention from the individual idea to the individual doing the problem solving. How students constructed complex ideas from resources was often so unique to that student that it was counter-productive to seek consistent patterns.[2] Instead, focusing on the student could give a rich perspective on one student's thinking, something far more aligned with our work as teachers interacting with individuals. Suddenly, the student and not the resource was the object of research. As my professional community changed to include teachers as experts in some areas and learners in others, my attention to the learner rather than their ideas continued. This, in turn, drove my move toward more naturalistic observation, which in turn strengthened the person-centered approach to the research.

In terms of methodology, the studies show a transition in my own career away from developing and analyzing survey instruments, though that is something I still do as part of my professional work [98], and toward a more naturalistic observation of interactions between people talking about science. My focus has expanded from question design to elicit knowledge to a recognition that the people I am studying care about far more than the nuances of a particular question and are attending to far more than knowledge alone in their classrooms. This awareness of and respect for complexity is mirrored in the data sources described in these papers: a narrow study on question design, an analysis of an isolated discussion between two peers, and a large group discussion among many teachers. The survey was part of a pre- and post-instruction assessment, the peer discussion was part of a videotaped group examination, and the large group discussion was part of a series of professional development meetings. There is an ongoing loss of control of the events happening in one's research environment, matched to a growth in complexity of the interactions between people.

---

[2] Not to say I didn't try! See citation [17].

Throughout, the resources framework has provided value in modeling the situation by honoring the situation for what is happening, and not for what is not happening. I began this work at a time when there was a call to understand student misconceptions – "if only we had a list of them!" was a statement heard during many workshops with teachers and instructors. But, as pointed out by Smith et al. [12], knowing the misconceptions alone didn't actually provide a pathway for helping students to construct a more refined understanding of a topic. Still, in the 1999 study, the goal was to determine the ideas students had, and how different questions were better or worse at eliciting them. By the 2007 study, some of those goals had continued, but there was an additional element regarding how students might use these resources differently in problem solving [46]. By 2017, when specific investigations of the resources framework were no longer the primary research agenda, the emergent analysis nevertheless pointed to the problems teachers might have in choosing between a resource and its many facets when responding to students in the classroom. I have summarized these changes and others discussed in this paper in Figure 2.

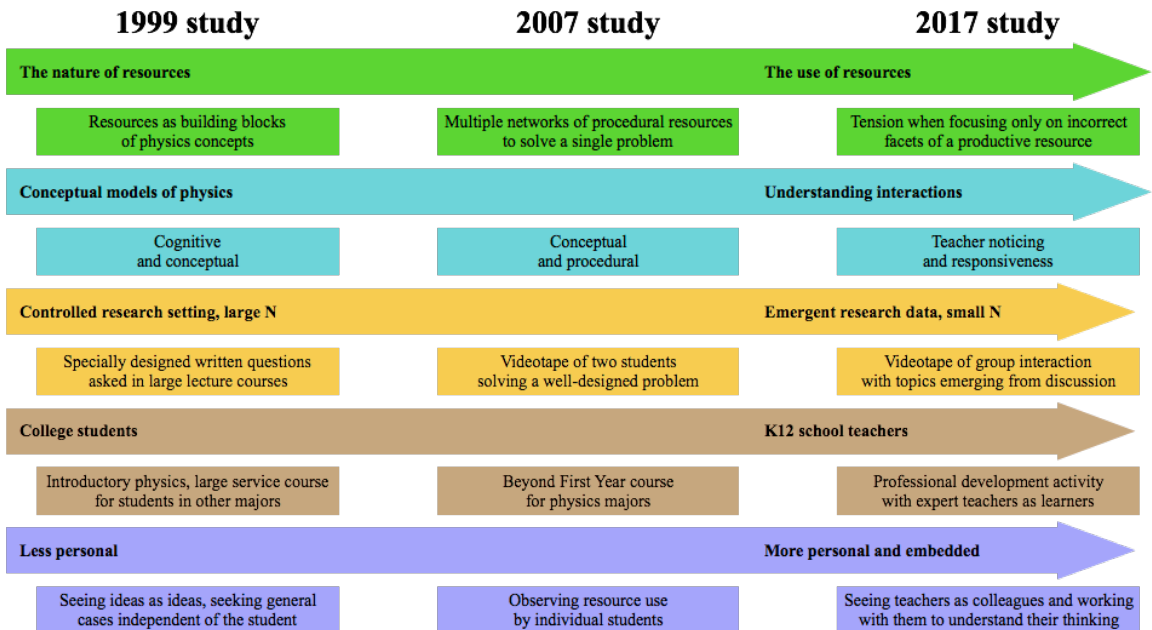

*Figure 2.* Transitions along multiple axes of how resources were studied over the course of three studies.

## V. DISCUSSION

In this section, I move from an analysis of the resources framework into a broader discussion of what has continued to motivate me in this work, and how that motivation has changed with time.

Quite obviously, the resources framework is a relatively narrow perspective for studying the great complexity of people's interactions with the world. It was designed to understand problem solving, but not everything is a problem to be solved. It has been used to investigate conceptual knowledge, epistemology, and the use of mathematical procedures, but as is made clear when working with teachers, these are but some of the issues which matter in the classroom, much less in the world outside of school. In other words, there is a narrow scope to the resources framework.

Its power, though, comes from its underlying structure. There is an acceptance that people hold contradictory ideas and may not sound coherent. There is a way of modeling how a person's actions depend in subtle ways on context. Most importantly, there is a sense of value to an individual's work, regardless of whether the application of a resource is useful or not, in a given situation. So much of our teaching work involves conversations about correctness, but the resources framework reframes this to a conversation about how students use resources; how they connect these resources when solving problems; what it was, specific to that student, which activated a particular idea; the consequences of those activations for that student's thinking; and much more. These are conversations about students constructing ideas, not just the ideas themselves. This mindset is deeply philosophically appealing to me, in part because it honors an individual's curiosity and creativity in a way that discussion of ideas, alone, does not.

In sum, the resources framework is generative, complex, and has space for people to have contradictory ideas. I have already provided an example, in the 2007 study, of how my personal life influenced my decision to study particular topics while using the resources framework. The opposite has also been true, namely that answering questions about the resources framework within physics education has helped me make more sense of the complexity and contradictions in my own life. That my professional work has been personally useful has been one reason I have persisted in using the resources framework over time. To illustrate this, I look at the topic of bilingualism again.

I am a bicultural immigrant, raised bilingually and raising my children bilingually. In each country I live in, I bring in some elements from the other country. My cultural habits include pieces from both places. Any attempt at coherence quickly falls apart, sometimes in absurd ways. According to those close to me, when I arrive from the USA in Austria, I change in subtle ways – I walk differently, gesture differently, and talk differently not just in my primary choice of language but also in the grammar structures I use in both English and German. I am mostly unaware of these things, even after others have repeatedly pointed them out to me. Engaging with the resources framework has helped me make sense of my own situation as I switch between languages, cultures, actions, and ideas. I can illustrate this using the 2007 study about integration, in which we modeled the ways that students use multiple methods to integrate in physics, with some resources shared between the methods and some unique to each method. This analysis helped me honor those resources that are part of only my American or only my Austrian life, while also honoring those that are consistent to both parts of my life. Rather than picking one or the other country, or worrying that I'm losing something by using one set of resources more in the one country than in the other, I can accept the parallelism, with overlapping structure, as a model for something that is happening in my own life. There is no right or wrong answer, there are simply two different right ways, activated differently in each context.

I have used the resources framework throughout my career in a wide variety of situations. The changes my work has gone through reflect in part the broader changes of the PER community. My research tools have changed from survey design and interviews in the 1990s, consistent with the community focus on standardized test design and curriculum development to address specific student difficulties, to group discourse and naturalistic video observation, consistent with a perspective based on social interactions and what each individual brings to the situation. My research environment has shifted from the introductory college topics that were the staple of college-focused PER when I joined the field to advanced physics topics and then to middle school instruction, consistent with expanding focus of discipline-based education research carried out by discipline specialists. My attention has shifted from conceptual knowledge in the 90s to procedural actions in the 00s to teacher noticing and responsive teaching in the teens. Most importantly, driven by an attention to the ways learners have constructed their individual solutions to the problems we ask, my attention has gone from the idea to the person having the idea. The flexibility of the resources framework is seen in its ability to drive these changes and be useful at each stage of the way.

## Acknowledgments

There are too many people to thank when considering the scope of this work. These include my mentors, Joe Redish, David Hammer, and Richard Steinberg; all my students and post docs whose work is cited throughout this paper; the colleague and collaborator who listened, questioned, and reflected back so many of my ideas over the years, Rachel Scherr; my colleagues at UMaine and in the Maine Physical Sciences Partnership; the teachers who have guided me so strongly since 2010; the members of the UMaine Physics Education Research Laboratory; and many, many more. The work was funded primarily by the National Science Foundation, with additional support from the US Department of Education and the Fund for the Improvement of Post-Secondary Education (FIPSE).

This paper will be published in the *Reviews in Physics Education Research* in a volume edited by (in alphabetical order) Charles Henderson, Kathy Harper, and Amy Robertson, publication date Summer 2018. See https://www.compadre.org/per/per_reviews/ for more information.